\documentclass{article}

\usepackage{arxiv}
\usepackage{setspace}
\usepackage[utf8]{inputenc} 
\usepackage[T1]{fontenc}    
\usepackage{hyperref}       
\usepackage{url}            
\usepackage{booktabs}       
\usepackage{amsfonts}       
\usepackage{nicefrac}       
\usepackage{microtype}      
\usepackage{lipsum}		
\usepackage{graphicx}
\usepackage{natbib}
\usepackage{doi}
\usepackage{float}
\usepackage{amsmath}
\usepackage{subcaption}
\usepackage[export]{adjustbox}
\usepackage{multirow}
\usepackage{xcolor}
\title{Bayesian Parameterized Quantum Circuit Optimization (BPQCO): A task and hardware-dependent approach}

\author{ \href{https://orcid.org/0000-0002-2720-7795}{\includegraphics[scale=0.06]{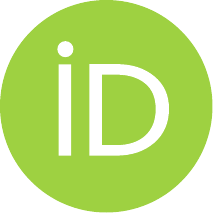}\hspace{1mm}Alexander Benítez-Buenache}
\\
	Department of Artificial Intelligence and Big Data\\
	GMV\\
	Isaac Newton, 11, Tres Cantos, 28760, Madrid, Spain \\
	abenitez@gmv.com \\
	\And
	\href{https://orcid.org/0009-0008-6354-7894}{\includegraphics[scale=0.06]{orcid.pdf}\hspace{1mm}Queralt Portell-Montserrat} \\
	Department of Artificial Intelligence and Big Data\\
	GMV\\
	Isaac Newton, 11, Tres Cantos, 28760, Madrid, Spain \\
	\texttt{qqpm@gmv.com} \\
}

\renewcommand{\shorttitle}{Bayesian Parameterized Quantum Circuit Optimization (BPQCO)}

\hypersetup{
pdftitle={Bayesian Parameterized Quantum Circuit Optimization},
pdfsubject={q-bio.NC, q-bio.QM},
pdfauthor={Alexander Benítez-Buenache, Queralt Portell-Montserrat},
pdfkeywords={Parameterized Quantum Circuits, Ansatz design, Bayesian Optimization, Noisy hardware},
}

\begin{document}
\maketitle

\begin{abstract}
Variational quantum algorithms (VQA) have emerged as a promising quantum alternative for solving optimization and machine learning problems using parameterized quantum circuits (PQCs). The design of these circuits influences the ability of the algorithm to efficiently explore the solution space and converge to more optimal solutions. Choosing an appropriate circuit topology, gate set, and parameterization scheme is determinant to achieve good performance. In addition, it is not only problem-dependent, but the quantum hardware used also has a significant impact on the results. Therefore, we present BPQCO, a Bayesian Optimization-based strategy to search for optimal PQCs adapted to the problem to be solved and to the characteristics and limitations of the chosen quantum hardware. To this end, we experimentally demonstrate the influence of the circuit design on the performance obtained for two classification problems (a synthetic dataset and the well-known Iris dataset), focusing on the design of the circuit ansatz. In addition, we study the degradation of the obtained circuits in the presence of noise when simulating real quantum computers. To mitigate the effect of noise, two alternative optimization strategies based on the characteristics of the quantum system are proposed. The results obtained confirm the relevance of the presented approach and allow its adoption in further work based on the use of PQCs.
\end{abstract}

\keywords{Parameterized Quantum Circuits \and Ansatz design \and Bayesian Optimization \and Noisy hardware}

\section{Introduction}\label{sec:intro}

The last few years can be regarded as the ultimate outcome of a journey that has been underway for decades towards the creation of impressive Artificial Intelligence (AI) systems. This progress is primarily attributed to the evolution of Machine Learning (ML) techniques \citep{bishop2023deep}: Neural networks (increasingly complex), attention mechanism, or the rise of generative AI, to cite a few examples. The evolution of hardware has played a significant role in facilitating the execution of training models, particularly through the use of GPUs and distributed computing. Consequently, large data sets can now be processed seamlessly. This ongoing evolution is leading to an increasing number of applications and improved performance. 

However, alongside the aforementioned development, a new area of research has arisen in the scientific community called Quantum Machine Learning (QML) \citep{biamonte2017quantum} \citep{schuld_intro_qml}, and it is becoming increasingly relevant. This new approach seeks to incorporate the principles of Quantum Computing (QC) into Machine Learning solutions. Thus, the aim is to incorporate the Physics properties of quantum states, such as superposition, interference or entanglement when carrying out the necessary calculations and operations. The main goal of this new paradigm is to achieve better solutions in terms of computational speed, better performance (the so-called quantum supremacy or quantum advantage) or even improvements in terms of energy consumption. QC represents a fundamental change in the handling of information. This computing approach sets the qubit (or quantum bit) as the basic unit of information. The primary contrast is that while a bit can represent either state 0 or state 1, a qubit can coexist simultaneously in a linear combination of states 0 and 1, thanks to the property of superposition of quantum states. To be clearer, if $|0\rangle$ and $|1\rangle$ are the two possible levels, a general quantum state $|\psi\rangle$ of this system is represented by $|\psi\rangle=\alpha|0\rangle+\beta|1\rangle$, with $\alpha,\beta\in\mathbb{C}$ and $|\alpha|^2+|\beta|^2=1$. 
 
Furthermore, qubits can be entangled. This is a quantum physics phenomenon where two particles are closely correlated so that there is a co-dependence between them in describing each other's quantum state, even when they are separated. It is important to note that this property will be of great importance when dealing with correlated data.

Quantum circuits are used to interact with quantum states. The circuits consist of a sequence of quantum gates, each of which acts as a unitary (\textit{i.e.}, reversible) operator. The primary functions of quantum gates include producing superposition states, phase modification, and generating entanglement. Additionally, the measurement of qubits in these circuits is noteworthy as it involves the collapse of a qubit into a quantum state. For instance, a qubit in the $|+\rangle$ state has an equal chance of obtaining 0 or 1 after measurement. However, the qubit will collapse into the state ($|0\rangle$ or $|1\rangle$) that has been measured. Hence, the measurement affects the quantum state of the system. By exploiting the physical properties described above, the combination of these gates allows the design of quantum algorithms, of which Shor's and Grover's algorithms are well-known examples. Throughout this paper it is assumed the knowledge of the different operations by means of quantum gates. Nevertheless, if the reader wants to become familiar with it, it is recommended to read \citep{nielsen2010quantum}. 

Despite the benefits of quantum computing that have been presented above, there are currently several limitations to this technology. In fact, this current period is known as the Noisy Intermediate-Scale Quantum (NISQ) era \citep{Preskill2018quantumcomputingin}, relative to the actual capabilities of quantum hardware. Typically, the low number of qubits has been the main limitation of quantum computers, which restricts the complexity and the amount of information that can be processed. For each quantum computer with $N$ qubits, there are $2^N$ quantum states within it. As a result, the number of possible quantum states increases exponentially with the number of qubits. Although this remains a limitation, it is worth noting the recent progress in the number of qubits available, which now exceeds a few hundred. Nevertheless, the main limitation of quantum systems in this NISQ era is the vulnerability of qubits to factors such as decoherence or quantum noise. This vulnerability results in computational errors due to the intrinsic hardware characteristics and external factors such as electromagnetic fields or thermal fluctuations. Additionally, each quantum computer has a different coupling map (scheme of possible connections between qubits) and, with few exceptions, not all qubits are fully connected to each other. Even the allowed gates may vary for each system. Thus, the circuits are transpiled. The goal of the transpiler is to adapt the circuit to be executed on a particular quantum computer. This process usually involves a drastic change in the circuit with respect to the ideal design, resulting in more complex and deeper (increasing gate layers) circuits. Obviously, the greater the depth, the greater the vulnerability to noise effects. For all these reasons, the importance of designing a suitable circuit is worth noting.

Nonetheless, the above limitations present a valuable opportunity to further research the field from both a hardware and algorithmic perspective. Undoubtedly, this parallel progress of algorithms and hardware has a strong analogy with the process carried out by classical ML, mentioned in the first lines of this section. In fact, the development of quantum algorithms, particularly QML algorithms, has been exceptional in recent years \citep{zaman2023survey}\citep{combarro2023practical}. As might be expected, most existing QML algorithms have emerged as adaptations of their classical counterparts. Good examples are Quantum Support Vector Machine (QSVM) \citep{rebentrost_qsvm_2014}, Quanvolutional Neural Networks \citep{henderson2020quanvolutional}, Quantum Convolutional Neural Networks (QCNNs) \citep{ohs_qcnn_2020}, or Quantum Recurrent Neural Networks (QRNNs) \citep{takaki_qrnn_2021}, to name a few. During the NISQ era, the primary approach is to evolve algorithms beyond the hardware restrictions. This has facilitated the progression of QML architectures design for solving small problems. However, the goal is to be prepared for hardware advances to tackle real-world problems, such as datasets with large amounts of data with multiple variables. In addition to this purely quantum approach, there are hybrid solutions \citep{sim_expressibility_2019} \citep{endo2021}. This approach aims to solve concrete tasks of a classical ML architecture in a quantum manner.

Within this brief compendium of techniques, Variational Quantum Algorithms (VQAs) stand out. This family of algorithms is based on the use of Parameterized Quantum Circuits (PQCs) \citep{Benedetti_2019}. As the name suggests, these circuits are designed to solve specific problems by optimizing a series of parameters, which commonly refers to the angle of a rotation gate. There are two main types of algorithms within the VQAs: Variational Quantum Eigensolvers (VQEs) and Variational Quantum Classifiers (VQCs). VQE is a quantum algorithm used to find approximations to the lowest energy states (eigenvalues) and corresponding eigenvectors of a Hamiltonian associated with a quantum system. It is typically used in optimization problems, but also in other tasks such as quantum chemistry and simulation. Similarly, VQCs use PQCs to implement classification models. These parameterized circuits are iteratively optimized using classical methods to minimize a loss function that measures the discrepancy between model predictions and the true labels of the training data. This type of circuit has the advantage of being easily integrated in hybrid architectures, \textit{i.e.}, architectures that combine quantum and classical layers in their structure. Thus, certain classical parameters, such as neural network weights, are replaced by rotation angles of the PQC. Various papers in the literature provide examples of this \citep{sakhnenko2022hybrid}\citep{chen_qlstm_2022}\citep{maheshwari_2022} \citep{sagingalieva2023hybrid} \citep{chen2024deep}. However, these circuits are problem-specific, meaning that the circuit's architecture (including the number of qubits, entanglement, gates, etc.) will depend on the problem that needs to be solved \citep{sim_expressibility_2019}. While circuit templates \citep{nakata17} \citep{hubregtsen2021evaluation} may be useful for certain tasks, it is important to note that there is no universal architecture that solves all problems. As a result, designing these circuits presents a significant new challenge, which is addressed in this paper. Ideally, designing the optimal quantum circuit would require time and knowledge of both the problem and quantum circuit design. However, this is impractical. Therefore, various search criteria have been studied, including the use of brute force \citep{e25070992}, genetic algorithms \citep{altares-lopez_autoqml_2022}, Bayesian optimization \citep{pirhooshyaran_quantum_2021}, reinforcement learning \citep{pirhooshyaran_quantum_2021}, or based on neural networks \citep{du2022quantum}. 

Approaches such as those mentioned above can help build effective circuits for specific problems, resulting in good performance. However, most approaches focus on designing ideal solutions without considering the current limitations of quantum hardware. As a result, when implementing these circuits on real hardware, their performance significantly degrades. Recent works \citep{buonaiuto2024effects} has already explored possible alternatives to take into account the coupling map in specific hardware.
In this context, we present a PQC design methodology, in which we highlight the following aspects:
\begin{itemize}
    \item Proposal of a complete and detailed process of PQC design, including notation to formulate the different variables to be explored during the search.
    \item Experimental tests that prove the effect of the circuit design on the results obtained.
    \item Demonstration of the vulnerability to noise of previously created designs in ideal environments.
    \item Presentation of two design alternatives that consider the characteristics and limitations of real quantum hardware.
\end{itemize}

The rest of the paper is structured as follows. Section \ref{sec:vqc_design} presents and details important criteria for designing PQCs. Section \ref{sec:quco} introduces BPQCO (Bayesian Parameterized Quantum Circuit Optimization), the proposal for circuit design using Bayesian Optimization, being evaluated with experimental examples in Section \ref{sec:experiments}. Section \ref{sec:real_hardware} evaluates the possible degradation in noisy environments of the obtained circuits and proposes an alternative to include the limitations of real hardware in the design process. Finally, Section \ref{sec:conclusions} presents the conclusions drawn from the work and future lines to be explored.

\section{On the design of Parameterized Quantum Circuits}
\label{sec:vqc_design}

Although there is no single structure of PQCs, several authors \citep{Benedetti_2019} \citep{du2022quantum}\citep{maheshwari_2022} point to the existence of two main blocks within the circuit: the feature map for data encoding and the ansatz. The first refers to the encoding of information in the circuit. That is, any d-dimensional data is transformed through a series of operations/gates to encode it as a quantum state. On the other hand, the ansatz is the most important part of the circuit, as its parameters are the solution to the problem to be solved. In the design of both blocks there are numerous options and criteria to take into account. In fact, both parts can be repeated or mixed with each other throughout the circuit. This section describes some of the key details of this design process to help the reader understand how it works and how it can be put into practice.

\begin{figure}[h]
    \centering
    \includegraphics[width=0.95\linewidth]{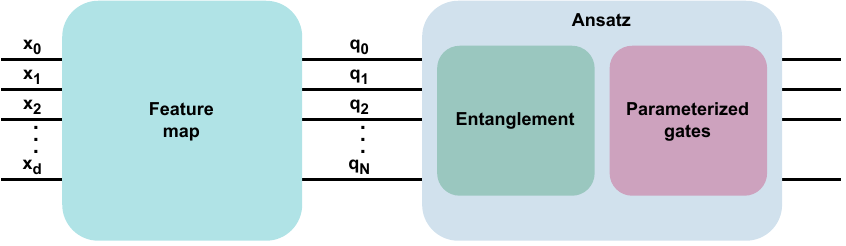}
    \caption{Schematic representation of the structure of an N-qubit PQC.}
    \label{fig:pqc_scheme}
\end{figure}

\subsection{Feature map}

Usually, classical data is represented as quantum states using a quantum feature map. Let $\mathcal{X}$ be the set of the input classical data. Then, a quantum feature map is a function $\phi: \mathcal{X} \rightarrow \mathcal{F}$, where $\mathcal{F}$ is a Hilbert space and the feature vectors are quantum states. Using a unitary transformation $U_{\phi}(x)$ it transforms $x \rightarrow |\phi(x)\rangle = U_{\phi}(x)|0\rangle$.

Thus, choosing an appropriate feature map is the first step in building a PQC, as it can affect performance, efficiency, and error susceptibility. The selection of a method for encoding data depends on various factors, such as the type of data (tabular, image, time series, etc.), the number of dimensions, and the number of qubits available on the chosen quantum computer. However, it is important to note that finding an encoder with enough expressive capacity to represent the information in the input data is sufficient. In practice, the most common methods for quantum encoding are basis encoding, amplitude encoding and angle encoding \citep{Schuld_2019} \citep{9425837}. Also of note is the \texttt{ZZFeatureMap} $-$used in later sections$-$, a circuit that encodes each feature of the data in a wire (\textit{i.e.}, the number of qubits required matches the number of features). This quantum circuit implements $U_{\phi}(x)$ by means of Hadamard (H) gates to create superposition, as they transform the $|0\rangle$ state into $1/\sqrt{2}(|0\rangle+|1\rangle)$. In addition, it also contains single-qubit rotations about the Z axis (whose angles depend on the value of the input's features) and CNOT gates. 

Finally, it is worth mentioning the possibility of using the data re-uploading technique, which has been shown to improve the performance of a universal quantum classifier \citep{P_rez_Salinas_2020}. This technique is based on repeating the encoding block along the circuit. 

\subsection{Ansatz}
After designing the encoder to inject the input data into the quantum circuit, it is time to design the ansatz. The ansatz is key in PQC as it provides the solution to the problem. The circuit's architecture is determined by the design of its components, such as the number of qubits, the level of entanglement, the depth, or the gates that are included. This has a major impact on the performance of the solution. However, their choice is not straightforward and is strongly problem-dependent. Therefore, this paper aims to help in the design of the ansatz architecture, as presented in the next section. 

Within the ansatz there are two main components: The gates that introduce the entanglement and the parameterized gates whose values are adjusted during training. Both blocks will be described in detail below. However, it is worth noting that they are not necessarily two sequential and distinguishable blocks, but two components to be considered.

As mentioned above, one of the main properties of quantum states is entanglement, which is the capacity of a quantum state in a qubit to influence the state of another qubit. There are several ways of creating entanglement between qubits, but the most common is to include a Hadamard gate followed by a CNOT. CNOT is a two-qubit gate that leaves the control qubit unchanged. If the control qubit is in the state $|1\rangle$ applies a Pauli-X rotation in the target qubit, and if it is in the state $|0\rangle$ leaves the target qubit unchanged. During the design phase of a PQC, layers of different two-qubit gate configurations, such as CNOT, CZ and other parameterized variants, are often added and repeated to produce strongly entangled circuits \citep{sim_expressibility_2019}. 

On the other hand, the ansatz will contain the part of the circuit whose parameters $\pmb{\theta}$ will be selected during training as the solution to the problem. Given a training dataset $\mathcal{X}=\{\pmb{x}^{(n)}, t^{(n)}\}, n=\{1,2,..., N\}$, where $\pmb{x}$ represents its features and $t$ its labels, the parameterized gates will act as a function $\mathcal{F}$ to reach the quantum state $\langle \hat{\mathcal{B}}\rangle=\mathcal{F}(\pmb{x}; \pmb{\theta})$ from the $\pmb{\theta}$ parameters. Thus, during circuit training, the aim is to achieve an optimal (or at least good enough) set of parameters
\begin{equation}
\pmb{\theta}^{*}= \operatorname*{argmin}_{\pmb{\theta}} \sum_N c(\pmb{t}, \langle \hat{\mathcal{B}}\rangle),
\end{equation}
where $c$ represents the subrogated cost (or loss function), a function that quantifies the error of the system output. In other words, the aim is to obtain the parameters that allow the input data to be transformed into a quantum state that allows the different classes of the classification problem to be distinguished. 

\begin{figure}[b]
    \begin{subfigure}[H]{0.33\textwidth}
    \includegraphics[width=\textwidth]{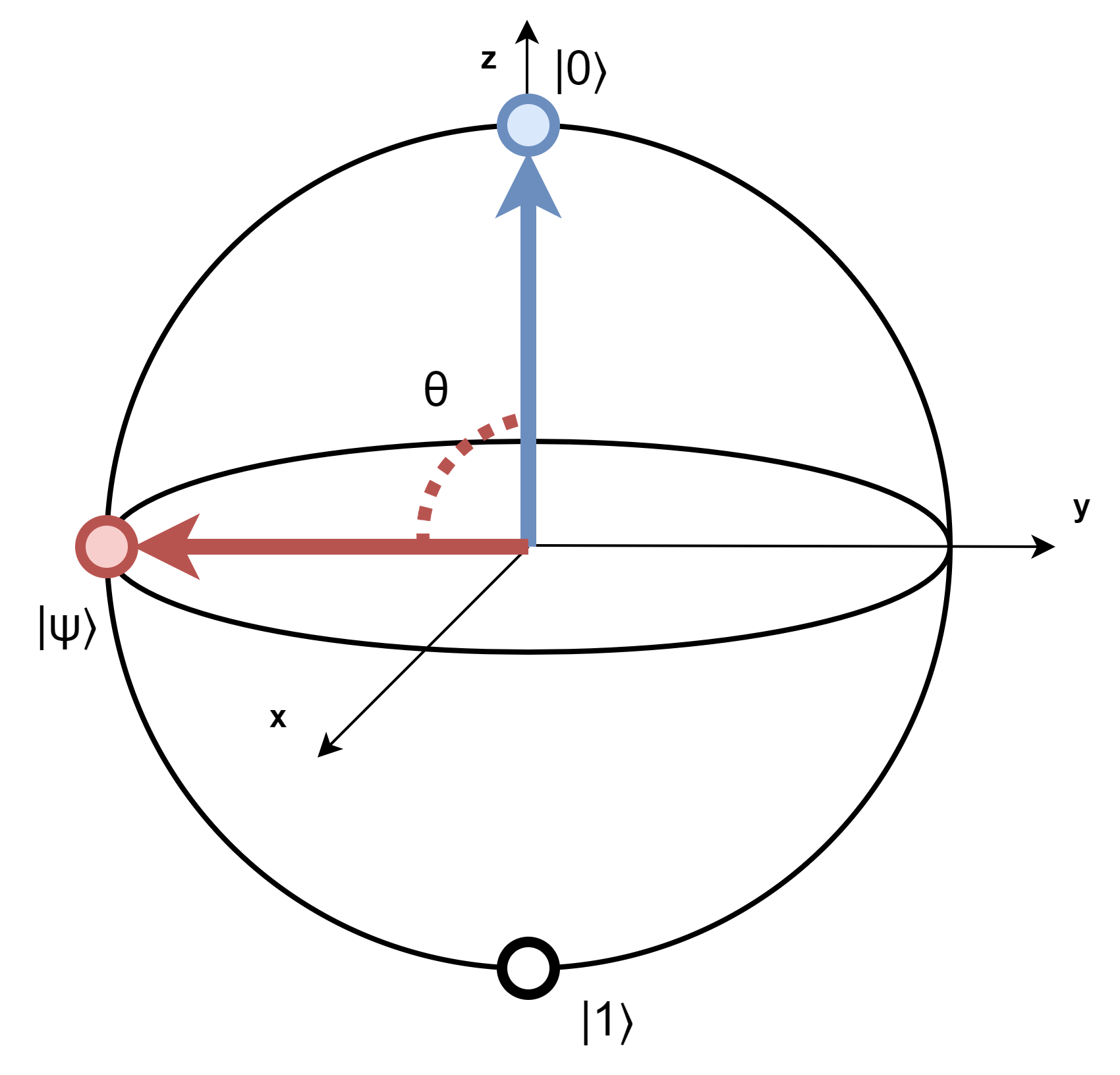}
    \caption{RX($\theta$)}
    \end{subfigure}
    \begin{subfigure}[H]{0.31\textwidth}
    \includegraphics[width=\textwidth]{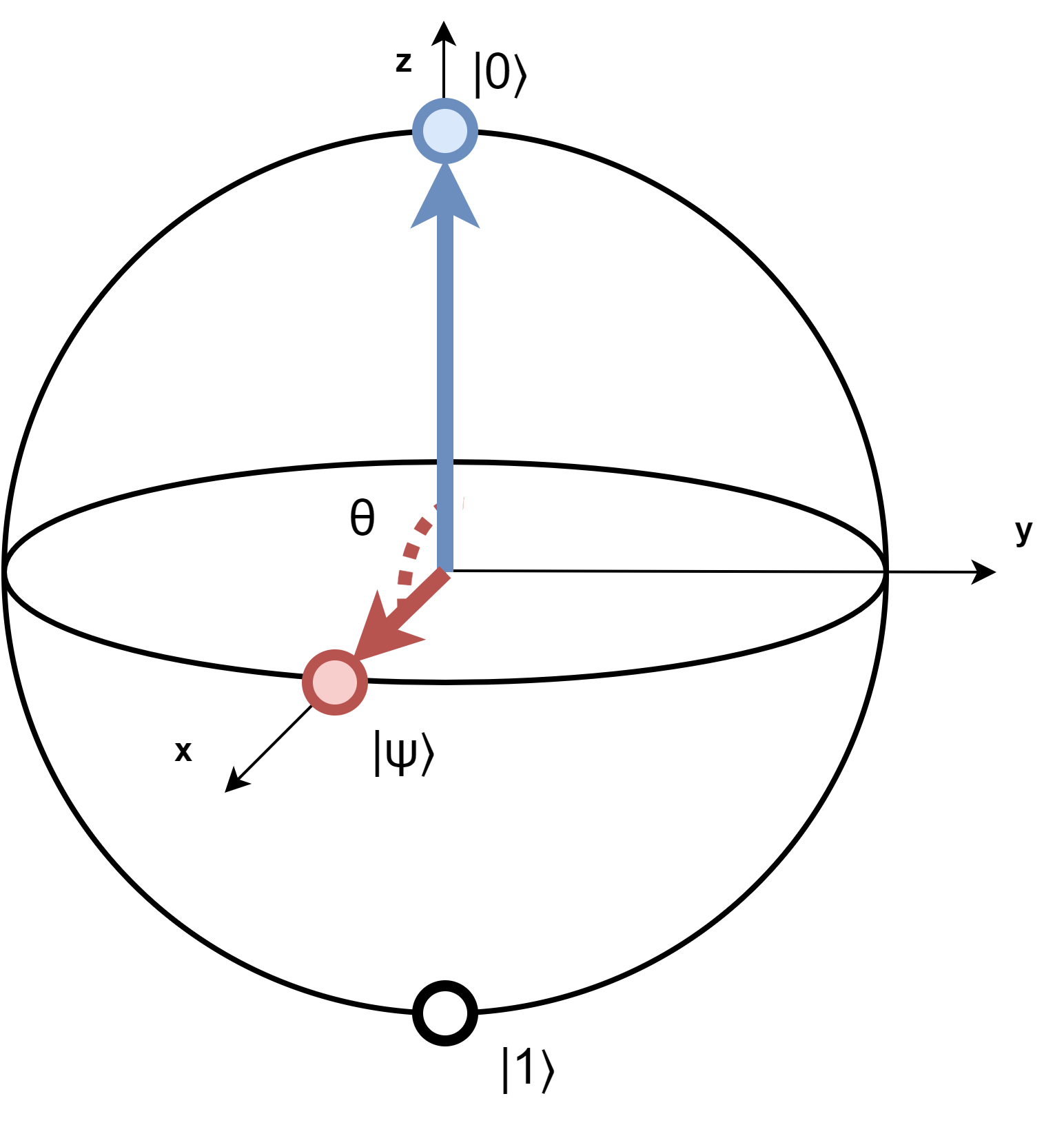}
    \caption{RY($\theta$)}
    \end{subfigure}
    \begin{subfigure}[H]{0.31\textwidth}
    \includegraphics[width=\textwidth]{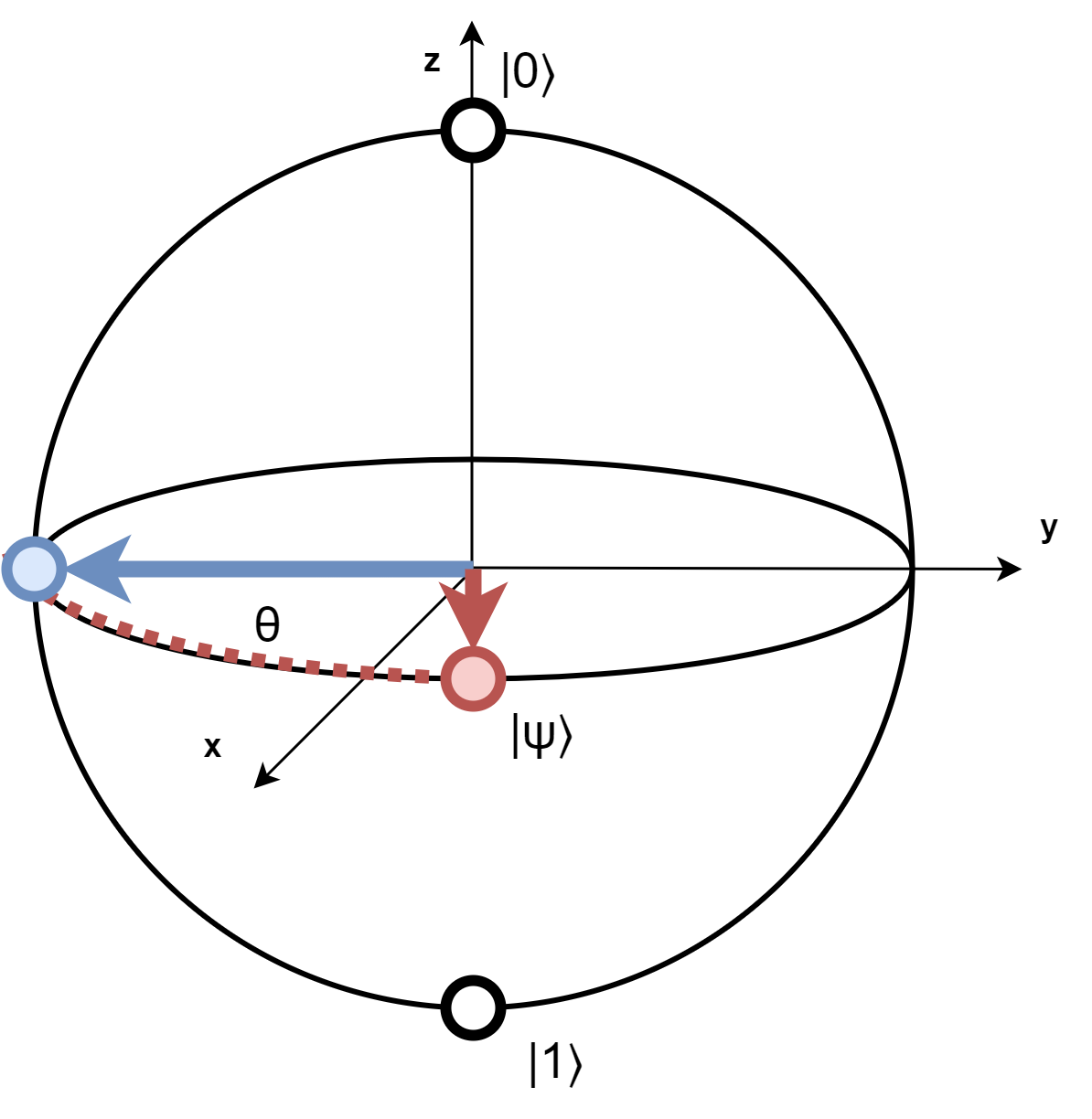}
    \caption{RZ($\theta$)}
    \end{subfigure}
\caption{Bloch sphere representation of the quantum state $|\psi\rangle$ change due to rotation $\theta$ at gates RX, RY and RZ, respectively. In blue the initial state and in red the state after rotation.}
\label{fig:rotations}
\end{figure} 

Within the PQC, these parameters are a series of rotation angles that are applied on the different qubits by means of rotation gates (RX, RY and RZ). The angle $\theta\in\{0,2\pi\}$ defines the rotation around the quantum state of the qubit in the X, Y and Z axes of the Bloch sphere\footnote{The Bloch sphere is a geometric representation to visualise the states of a qubit. It maps the possible quantum states onto the surface of a unit sphere, where the poles represent the basis states $|0\rangle$ and $|1\rangle$ and the points on the sphere's surface represent superposition of these states.}. Figure \ref{fig:rotations} shows three examples of rotation using each of these gates. In addition, these rotations can be controlled by another qubit in the same way as for the CNOT gate. If the control qubit is in the state $|1\rangle$, the corresponding rotation is applied to the target qubit. Whereas if the control qubit is in the state $|0\rangle$, the target qubit remains unchanged. These controlled gates will be referred as CRX$_{ij}$, CRY$_{ij}$ and CRZ$_{ij}$, where the indices $i$ and $j$ indicate that the qubit $q_i$ is the control qubit and the qubit $q_j$ is the target qubit. 

Once the possible gates for the design have been presented, it is necessary to select the architecture. This is a process similar to the definition of a classical neural network, which requires certain criteria to decide on aspects such as the number of hidden layers, the number of neurons in each layer and other hyperparameters that are not trainable. In this case, the choices are the configuration of the rotation gates for each qubit wire in terms of the number of gates (depth of the circuit), the type of rotation gate (axis of rotation and number of qubits involved in the rotation) and the position of each of them. In fact, the order and position of the gates can have a major impact on the performance of the PQC. This fact motivates the possibility of using search criteria to obtain an optimal (or at least a good enough) circuit, as presented in the following section.

\section{BPQCO: A Bayesian Optimization approach for PQC design}\label{sec:quco}
As seen in the previous section, there are numerous aspects and criteria for the design of a quantum circuit. Beyond the possibility of carrying out preliminary exploratory studies to analyse the possible effects of incorporating certain gates, the number of possibilities and the need to train the system to see the performance (with the time involved) makes brute force approaches unfeasible. It is therefore necessary to establish other types of strategies to maximize performance for a given problem. This paper proposes BPQCO (Bayesian Parameterized Quantum Circuit Optimization), a Bayesian Optimization based approach to address this need. This option has been explored in some previous works \citep{pirhooshyaran_quantum_2021}, \citep{koikeakino2022autoqml} with promising results. However, the aim is to go one step further with more complex architectures and an exhaustive analysis of the behaviour of the solutions obtained (as will be seen in later sections).

To this end, we propose to define each possible gate position as a hyperparameter of the PQC to be designed. Each hyperparameter can take different values depending on the circuit block in which it is included. Then, based on a Bayesian optimization process, the values (gates) that maximise the performance of the circuit are selected. This process is detailed below, including the definition of each hyperparameter to be explored and the proposed search method.

\subsection{PQC architecture to be explored}

To optimize the circuit architecture, it is necessary to define some design criteria first. This paper follows the scheme presented in Section \ref{sec:vqc_design}, which consists of a feature map followed by the ansatz, including entanglement and parameterized gates. However, it is also valid to consider alternative options, such as fixing certain components or rearranging the order of elements. Therefore, the objective is to select the optimal circuit $\mathcal{C}^*= (\mathcal{D}, \mathcal{M}, \mathcal{E}, \mathcal{P})$ that maximizes performance, with $\mathcal{D}$ being design parameters (such as the number of qubits or the circuit depth), $\mathcal{M}$ the feature map, $\mathcal{E}$ the entanglement, and $\mathcal{P}$ the parameterized gates.

The circuit optimization process can include the choice of design parameters $\mathcal{D}$ to consider circuits with varying numbers of qubits ($N_{qubits}$) or depth ($N_{gates}$) for optimal performance. However, these aspects can be predetermined through exploratory tests based on hardware or resource constraints such as computation time and simulation system memory. 

In terms of the feature map, there are several options to consider, including the encoding method and the intrinsic parameters of these methods, such as the angle of rotation. Each option determines a parameter to explore during optimization. However, as previously mentioned, for the encoding it is sufficient to find a method to adapt the information with enough expressiveness. Therefore, it is recommended to establish $\mathcal{M}$ a priori, after conducting some preliminary exploratory tests.

\begin{figure}[t]
    \centering
    \includegraphics[width=0.95\linewidth]{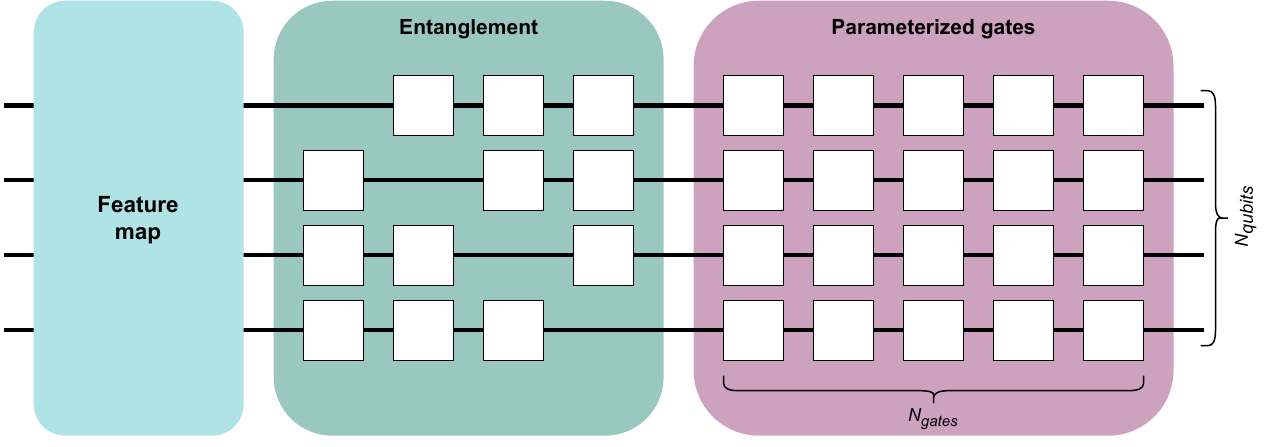}
    \caption{Graphical description of the design elements to be selected during circuit optimization. Each white block represents the position where a gate can be selected.}
    \label{fig:pqc_to_explore}
\end{figure}

The proposed method for selecting the gates that introduces entanglement between qubits involves using matrix
\begin{equation}\label{eq:entanglement}
\hat{\mathcal{E}}=
    \begin{cases}
        0, & \forall i=j; \; i,j=0,1,...,N_{qubits}-1\\
        e_{ij}\in\{0,1\}, & \forall i\neq j; \; i,j=0,1,...,N_{qubits}-1,
    \end{cases}
\end{equation} 
which consists of $(N_{qubits}\times N_{qubits})$ binary variables. Each variable represents the presence or absence of a CNOT gate between qubits $i$ and $j$ for their entanglement. For obvious reasons, the diagonal is always zero, since it is impossible to create entanglement between a qubit and itself.

On the other hand, the configuration of parameterized gates that form the ansatz solution are modeled by
\begin{equation}\label{eq:parameterized_gates}
\begin{split}
 \hat{\mathcal{P}}=p_{ik}\in\{\mathrm{None}, \mathrm{RX}, \mathrm{RY}, \mathrm{RZ}, \mathrm{CRX}_{ij}, \mathrm{CRY}_{ij}, \mathrm{CRZ}_{ij}\} \\ 
 i,j=0,1,...,N_{qubits}-1, \; j\neq i, \\ 
 k=0,1,..., N_{gates}-1,      
\end{split}
\end{equation} 
a matrix of size $(N_{qubits}\times N_{gates})$ composed of categorical variables. Each variable represents a parameterized gate that will be placed at each position $k$ of the $i$-qubit. As shown in (\ref{eq:parameterized_gates}), the available options include one-qubit and two-qubit (controlled) rotation gates, as well as the option of not including any gate (None).

To ensure efficient use of the gates and to avoid increasing the circuit depth unnecessarily, it is recommended to apply post-processing criteria. To this end, we propose to establish two criteria: 1) Remove repeated consecutive gates only if they are simple rotations or if there is no gate in the qubit with which it is entangled; 2) Eliminate RZ gates in the last position since, being measured in z, they only affect the global phase.

Figure \ref{fig:pqc_to_explore} illustrates the proposed circuit structure. However, as mentioned earlier, there are numerous design possibilities. For instance, in the figure, the entanglement and parameterized gates appear as independent components. However, an alternative approach would be to combine them into a single block where either CNOT gates or rotation gates are selected for each position. Therefore, it is important to note that regardless of the initial design criteria, the use of a circuit optimization method is essential, as explained below.

\subsection{BPQCO by means of TPE algorithm}

After defining the options for building the PQC, it is necessary to establish an optimization strategy. For this purpose, we propose BPQCO, a Bayesian optimization process based on the Tree-structured Parzen Estimator (TPE) algorithm \citep{NIPS2011_86e8f7ab}. The TPE algorithm is commonly used in ML for hyperparameter optimization. To accomplish this, an objective function must be chosen to evaluate performance and then minimized or maximized accordingly. In this case, the hyperparameter to be evaluated are those defined in the previous section ($\mathcal{D}, \mathcal{M}, \mathcal{E}, \mathcal{P}$). These hyperparameters are defined as binary and categorical variables according to (\ref{eq:entanglement}) and (\ref{eq:parameterized_gates}), respectively. The usual strategy for optimizing an objective function assumes that the objective function exists in a continuous domain. Therefore, the algorithm rounds the proposed continuous points to the nearest discrete points. During the optimization process, a guided and efficient search is conducted towards promising regions of the hyperparameter space using a probabilistic model of both evaluated and non-evaluated parameters. Thus, several combinations of hyperparameters are evaluated during the search by means of trails. The benefit of this approach is that, after a limited number of trials, a good value for the objective function is achieved.

To do this, the BPQCO process selects circuits (trials) that are trained and tested in a validation subset. Each of these circuits obtains a metric in this validation subset, which will be the objective function of the Bayesian optimization process. After the guided search, the circuit with the best metric is selected.

\section{Experimental examples}\label{sec:experiments}
This section provides examples of how the proposed BPQCO methodology works. For the tests, we have chosen two classification problems (binary and multiclass). Therefore, the selected algorithm for the task is a VQC. Thus, the main goal is to find the ansatz that gives the best possible performance for each classification problem. That is, not to use a pre-defined ansatz, but to adapt it to the specific problem to be solved.

\begin{figure}[b]\centering
\includegraphics[width=1.055\textwidth]{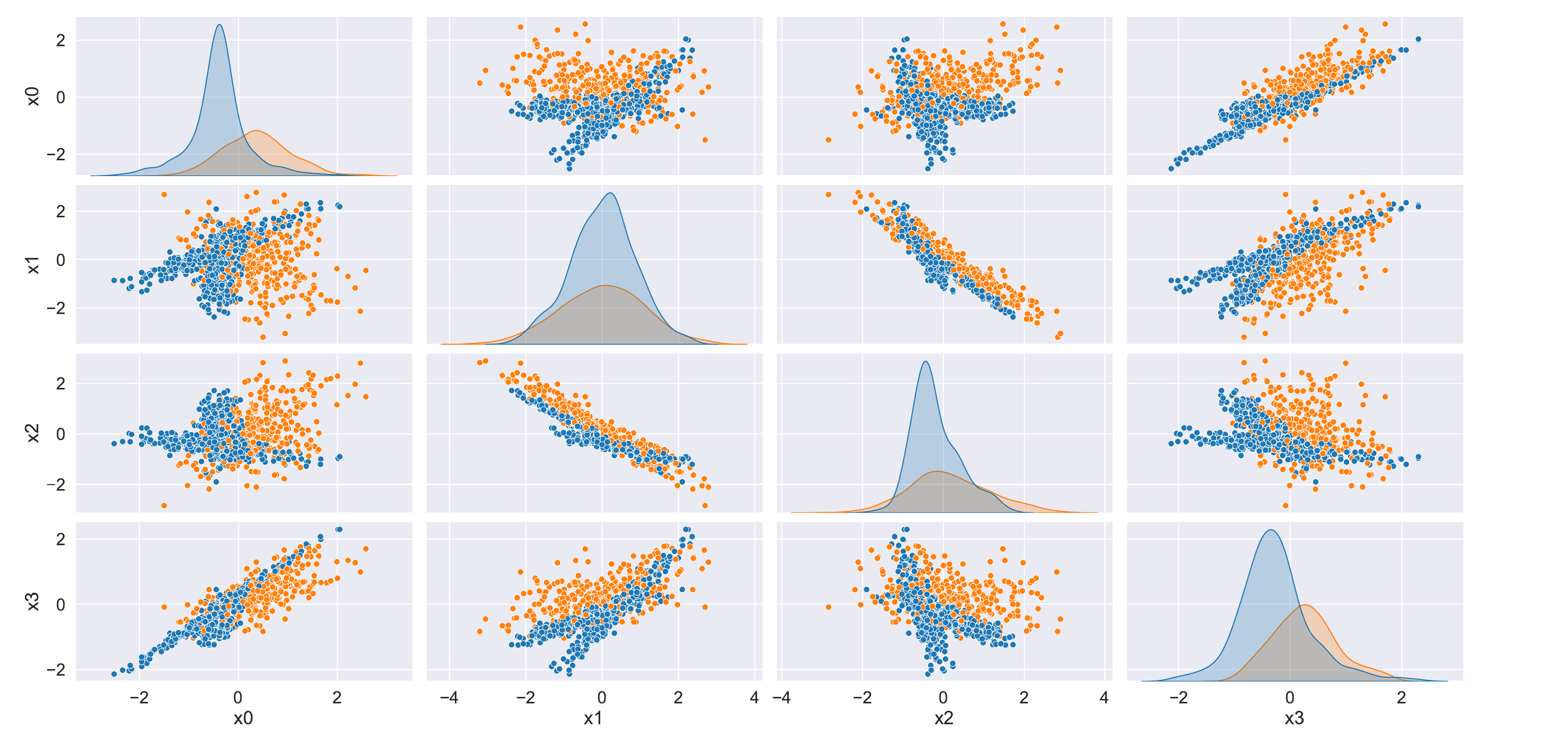}
\caption{Classes distribution and relationship between features for the Synthetic dataset.}
\label{fig:sns_plot}
\end{figure}

\subsection{Datasets}
A synthetic dataset and a real-world problem have been used for testing. Both are described below:
\begin{itemize}
    \item \textbf{Synthetic dataset (binary classification)}: This synthetic dataset has been generated by creating clusters of $4$-$d$ points normally distributed about vertices of a bidimensional hypercube with sides of length $0.8$. In this case, an equal number of clusters is assigned to each class. The number of generated samples is $1000$ with a class imbalance ratio of $65:35$ and a class separation of $0.4$. Figure \ref{fig:sns_plot} shows the distribution of both classes.
    
    \item \textbf{Iris dataset (multiclass classification)}: The Iris dataset \citep{misc_iris_53} contains $150$ instances described by four continuous features. The objective is to classify the three types of Iris plant: Setosa, Versicolour, and Virginica. It is a well-known dataset typically used in ML research. It is an easy problem with high performance using classical algorithms. However, the reduced number of samples allows to perform test series with lower computational load.
\end{itemize}

\subsection{Architectures and environment description}
The learning architectures used for each of the datasets are detailed below.

First, a C-Support Vector Machine (SVC) is trained to have a classical baseline. Its hyperparameters ($C=1.0$ and Radial Basis Function kernel) are set by default and have not been optimized, since the objective of this paper is not to achieve better results than the classical approach. 

The VQC circuit is defined as $\mathcal{C}=(\mathcal{D}, \mathcal{M}, \mathcal{E}, \mathcal{P})$, following the same structure as defined in the previous sections. This allows the use of BPQCO approach. For the experiments, certain parameters are fixed a priori, so that the focus is solely on the construction of a specific ansatz for each classification problem. The design parameters $\mathcal{D}$ are set to $N_{qubtis}=4$ and $N_{gates}=5$. The reason for selecting that number of qubits is that both datasets have four features. As for the maximum number of gates per qubit, it has been selected based on some preliminary tests. Additionally, the \texttt{ZZFeatureMap} circuit is chosen as the feature map $\mathcal{M}$, since the number of features and $N_{qubits}$ are the same. Thus, the circuit to be optimized $\hat{\mathcal{C}}$ depends only on the ansatz design: $\hat{\mathcal{C}}=(\hat{\mathcal{E}}, \hat{\mathcal{P}} | \mathcal{D}, \mathcal{M})$. To highlight the value of the proposed approach, it is worth noting that the number of possible combinations is $7.78\cdot10^{25}$, due to the $12$ binary variables and $20$ categorical variables (with $13$ possible values) to be explored. 

In addition, the following common ansatz templates are used to compare performance and establish a quantum baseline:
\begin{itemize}
    \item \texttt{RealAmplitudes}: This circuit template is composed of alternating layers of Y-rotations and CX gate entanglements used as a classification circuit in QML. The prepared quantum states have real amplitudes (the complex part is always zero).
    \item \texttt{EfficientSU2}: The architecture of this circuit is composed of layers of single qubit operations spanned by SU(2) and CX entanglements. This heuristic pattern serves as a tool to prepare trial wave functions for VQAs or to construct VQCs for QML. SU(2) refers to Special Unitary group of degree $2$, using $2\times2$ unitary matrices with determinant $1$, like the Pauli rotation gates.

    \item \texttt{PauliTwoDesign}: Comprising alternating rotation and entanglement layers, the circuit begins with an initial layer of $\sqrt{\mathrm{H}}= $RY$(\pi/4)$ gates. The rotation layers feature single-qubit Pauli rotations, randomly selecting axes from X, Y, or Z. The entanglement layers consist of pairwise CZ gates with depth of 2. It is a particular form of a two-design circuit \citep{nakata17}, which is frequently studied in QML.

\end{itemize}

\texttt{PauliTwoDesign} circuit has its own entanglement scheme, while the other two have different type of entanglement implemented: linear, reverse lineal (full) and circular entanglement. Additionally, all of them are tested with different repetition numbers (ranging from 1 to 5) to evaluate their performance across various circuit depths. Obviously, to establish the baseline, the best performing configuration is selected.

In this case, all VQCs, both the templates and the circuits obtained using the BPQCO approach, are simulated in an ideal quantum environment (the reader is advised that noise will be included in the next section of the paper) using the IBM \texttt{Qiskit} library\footnote{The versions used for testing are as follows: \texttt{qiskit}: 0.44.1, \texttt{qiskit-aer}: 0.12.2, \texttt{qiskit-algorithms}: 0.2.1, \texttt{qiskit-machine-learning}: 0.6.1, and \texttt{qiskit-terra}: 0.25.1.}. This includes the absence of noise and decoherence errors, as well as full connection between all qubits. For that purpose, the Sampler\footnote{The Sampler is an IBM \texttt{Qiskit} service that computes the quasi-probabilistic distribution of bitstrings from quantum circuits.} used to simulate the performance of an ideal quantum computer has a noiseless backend constructed with the density matrix method and $1024$ shots. 

Finally, the COBYLA (Constrained Optimization BY Linear Approximation) optimizer \citep{cobyla_powell} is selected to train each of the VQCs, as it is a non-gradient based optimization method that significantly speeds up the learning process. The maximum number of evaluations of the function is set to $100$.

\subsection{Results and discussion}  
To obtain results, the $K$-Fold method is used with $K=10$ and a ratio of $0.7/0.3$ for training/validation subsets. In this way, each VQC is independently trained $10$ times, and the average accuracy in the validation subsets is obtained as a metric for performance comparison.

First, the SVC is implemented for each of the datasets, obtaining an accuracy of $0.8263\pm0.0142$ for the synthetic dataset and $0.9533\pm0.0271$ for the Iris dataset. As mentioned above, the SVC hyperparameters are not optimized. However, this allows us to establish a benchmark from a classical point of view, although the goal of this paper is not to outperform them. 

The next step is to establish a quantum baseline using the circuit templates described above, and to obtain the best entanglement scheme and number of repetitions for each of them. This is done by running all possible configurations to get the best possible performance and to establish a true baseline. The best configurations obtained for each type of circuit are shown in the Table \ref{tab:baseline_configurations}. 

Finally, the BPQCO approach is performed to obtain the most suitable circuits for each of the datasets. For this purpose, 600 and 700 trials are run for the Synthetic and Iris datasets, respectively. After that, the best ansatze are obtained, which are shown in Figure \ref{fig:best_circuits}.

\begin{table}[t]
    \centering
    \caption{Configuration of each type of circuit template to set the baseline for each dataset. Note that \texttt{PauliTwoDesign} has a predefined entanglement scheme.}
    \begin{tabular}{lccccc}
    \toprule
         &  \multicolumn{2}{c}{\texttt{RealAmplitudes}}  & \multicolumn{2}{c}{\texttt{EfficientSU2}}  &  \texttt{PauliTwoDesign}  \\
         \cmidrule(r){2-6}
         dataset & reps. & entanglement & reps. & entanglement & reps. \\
         \cmidrule(r){1-1}
         \cmidrule(r){2-3}
         \cmidrule(r){4-5}
         \cmidrule(r){6-6}
         Synthetic & 4 & Full & 5 & Full & 2 \\
         Iris & 2 & Linear & 3 & Full & 3  \\
    \bottomrule
    \end{tabular}
    \label{tab:baseline_configurations}
\end{table}

Table \ref{tab:noiseless_results} presents the results obtained with the different circuits. It shows the average accuracy and its standard deviation in the K-Fold validation subsets with 5 different initializations of the Sampler of the quantum environment (under ideal conditions). These initializations are done in order to study also the stability of the solutions. As can be seen, the results obtained with BPQCO circuits significantly outperform those of typical template circuits. In addition, BPQCO's results show increased stability of its solutions. Therefore, it can be concluded that it is important to adapt the circuit for each problem. However, in the next section we go a step further to analyze the behavior of these circuits under noisy conditions.

\begin{table}[t]
	\caption{Results obtained using the different ansatze for each dataset in terms of accuracy. The subscript $i$ indicates that the quantum simulation environment is under ideal conditions.}
	\centering
	\begin{tabular}{lll}
		\toprule
		& \multicolumn{2}{c}{Dataset}                   \\
		\cmidrule(r){2-3}
		Ansatz & Synthetic & Iris  \\
		\midrule
		\texttt{RealAmplitudes}$_i$ & $0.7448\pm0.0407$ & $0.7667\pm0.0559$     \\
		\texttt{EfficientSU2}$_i$     &
        $0.7465\pm0.0348$ & $0.7729\pm0.0657$     \\
		\texttt{PauliTwoDesign}$_i$     & $0.6507\pm0.0550$ & $0.7400\pm0.0770$    \\
        BPQCO$_i$     &  $\pmb{0.8124\pm0.0231}$&$\pmb{0.9133\pm0.0393}$   \\
		\bottomrule
	\end{tabular}
	\label{tab:noiseless_results}
\end{table}

\section{A first step towards real quantum environments}
\label{sec:real_hardware}

Up to this point, the paper has presented a process for finding suitable circuits for a given problem. However, this has been done from an ideal point of view without considering noise, decoherence errors, or the coupling map among qubits. Therefore, it is expected that the performance of the previously obtained circuits will not be as good when implemented on real quantum hardware. Thus, this behavior is analyzed in this section and two alternatives are proposed to adapt the circuits to existing quantum environments.

\subsection{Noisy environment simulation}

\begin{figure}[t]\centering
\includegraphics[scale=0.5]{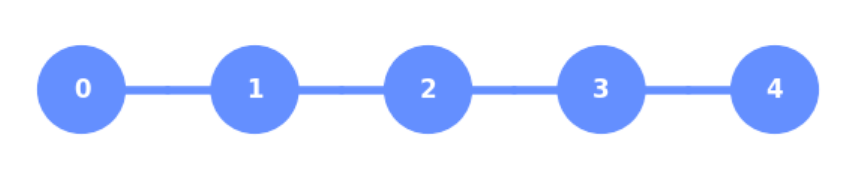}
\caption{Coupling map (qubit interconnection) of the \texttt{FakeManila} backend.}
\label{fig:coupling_map}
\end{figure}

First, a simulation environment is chosen that takes into account the real characteristics of a quantum computer. The reason for opting for simulation is the time associated with execution on real quantum hardware (waiting queues or the connection interface between the execution environment, the platform, and the hardware). In order to obtain reliable results, each circuit has to be executed several times. In addition, it has been found that the BPQCO method requires the execution of hundreds of trials to improve the results. For both reasons, execution on real hardware would not allow exhaustive testing due to the time required. For this purpose, IBM's \texttt{FakeBackend} is chosen. The fake backends are designed to replicate the actions of IBM Quantum systems by using system snapshots. These snapshots contain key information about the quantum system, such as the coupling map, basis gates, qubit properties (as the error rate), which are useful for testing the transpiler and running noisy simulations of the system. It is important to note the significance of the transpiled circuit, since it is composed only of those gates supported by the specific quantum computer being used, also considering the coupling map. This is due to the fact that gates involving multiple qubits may need intermediate SWAP gates to match the coupling map. In this case, \texttt{FakeManila} is chosen, a $5$-qubit backend that simulates the real operation of IBM's Manila quantum computer. Figure \ref{fig:coupling_map} shows the coupling map of the selected backend, in which the actual connection between the different qubits can be observed. On the other hand, the basis gates supported by this environment are the following: ID, RZ, SX, X, CX, and RESET. Therefore, all circuits executed in such environment must be transpiled to use only these gates.

Once the environment is prepared, the circuits obtained by BPQCO$_i$ (under ideal conditions) are executed in this noisy environment. The Synthetic dataset shows a significant decrease in performance, with an accuracy of $0.7027\pm0.0589$, representing a $10.97\%$ degradation. Similarly, the Iris dataset also experiences a considerable degradation ($48.75\%$), with an accuracy of $0.4258\pm0.0664$ in the noisy environment. The baseline circuits also suffer degradation, with a $5.52\%$ and $8.05\%$ decrease in performance for the Synthetic and Iris datasets, respectively. These results were expected and, in fact, reinforce the need to adapt the procedure presented above to the noisy environment. Thus, two potential alternatives are presented below.

\subsection{Circuit optimization under noise conditions}
In order to adapt the designed circuits to both the problem to be solved and the quantum hardware to be used, two design alternatives based on the BPQCO process are presented. The first will be executed directly in the simulated noisy environment, while the second will attempt to incorporate some of the characteristics of the noisy environment into the circuit optimization process.

\subsubsection{BPQCO in the noisy environment}\label{sec:bpqcon}

The first alternative is to implement the BPQCO approach directly in the simulated noisy environment. This requires transpiling each circuit designed for execution in the environment (in this case, \texttt{FakeManila} backend), taking into account the supported gates and the coupling map. The simulator will also consider the error characteristics of each qubit associated with the noise. Thus, the process is the same, but executed in a noisy environment. The circuit that maximizes the objective function should offer consistent performance when executed in real hardware. 

This approach's execution times are longer than the ideal case due to simulating all the effects of the real environment. However, these run times are still much lower than those associated with direct execution on real hardware. Direct execution on the quantum computer would entail a higher cost of resources (economic and time). Therefore, it is recommended to simulate the circuit optimization based on the problem requirements and available systems before implementation.

\subsubsection{Multi-objective BPQCO}\label{sec:mobpqco}

As a second option, the possibility of finding a suitable circuit for a specific hardware by altering the optimization process is explored. For this purpose, we propose the use of a multi-objective optimization. In this approach there are now two objective functions: The metric (whose performance is to be maximized) and the circuit complexity (which is to be minimized). For this purpose, the Bayesian TPE algorithm is again used, but adapted to multi-objective search \citep{ozaki2022multiobjective}. The motivation for this new approach is to incorporate real environment information while still working in an ideal environment to reduce execution time.

To do this, it is necessary to define the objective function related to the complexity of the circuit. For this purpose, the transpiled circuit $\mathcal{C_T}$ is used, considering the supported basis gates and the coupling map, which implies an adaptation of the circuit $C$. This leads to a larger number of basis gates. Once $\mathcal{C_T}$ is known, its complexity is computed by summing the error associated with each of its basis gates in each qubit, values that are available in the backend properties. Therefore, by means of the multi-objective search, the aim is to obtain the transpiled circuit with the lowest possible complexity (to be less susceptible to errors) that obtains the best results.  

\subsection{Results under noise conditions and discussion}  

To complete the experimental tests, the two design alternatives described above are used for the same datasets (Synthetic and Iris). Again, \texttt{FakeManila} backend is used to simulate the real environment. In addition, all tests are carried out using the same criteria as above to allow a proper comparison: $10$-Fold with the same splits between train/test in each fold and averaging of the accuracy in $5$ different runs by changing the backend initialization seeds.

First, a preliminary search for the best predefined circuit to set a noisy baseline performance, denoted as baseline$_n$, is performed. In both datasets, the best performance with a predefined circuit is obtained with the \texttt{EfficientSU2} circuit. In the case of the synthetic dataset, with circular entanglement and $1$ repetition. For the Iris, with reverse linear entanglement and $3$ repetitions.

\begin{figure}[H]
\centering
\renewcommand{\arraystretch}{0.3}
   \begin{tabular}{c}
   \includegraphics[width=1\linewidth]{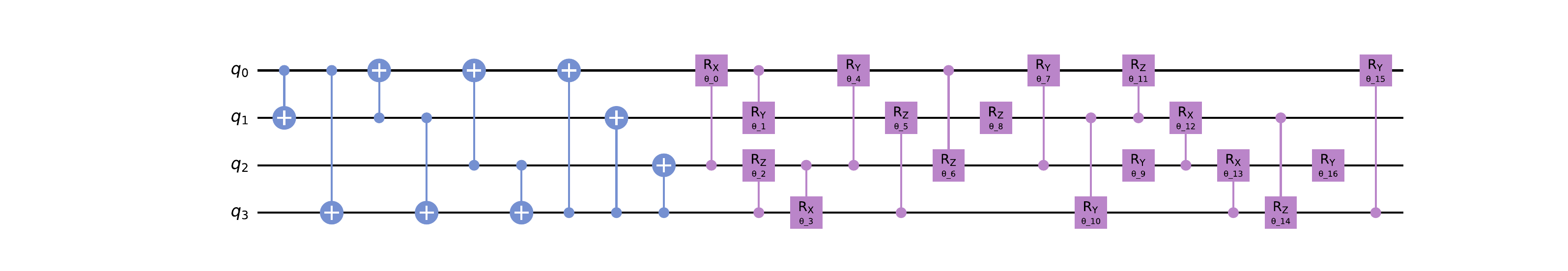}\\[0pt]
   (1.a)
\\[0pt]
   \includegraphics[width=1\linewidth]{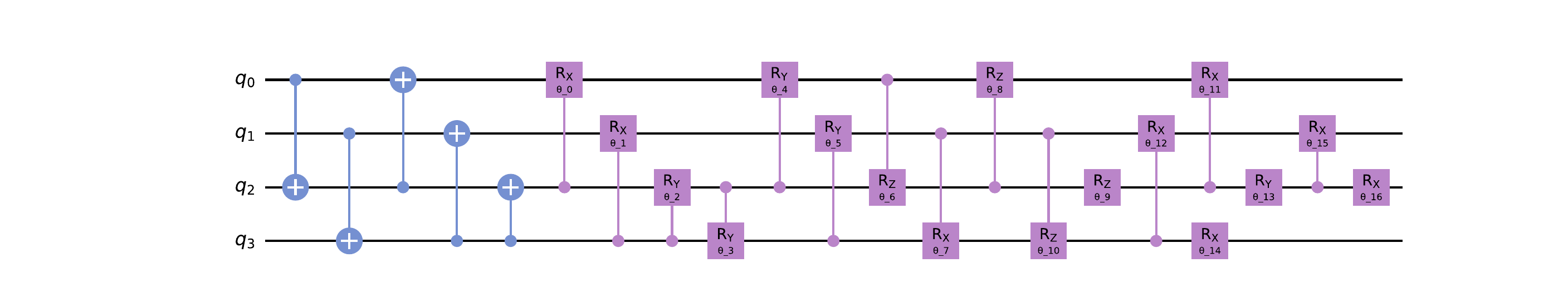}\\[0pt]
   (1.b)
\\
  \includegraphics[width=0.65\linewidth]{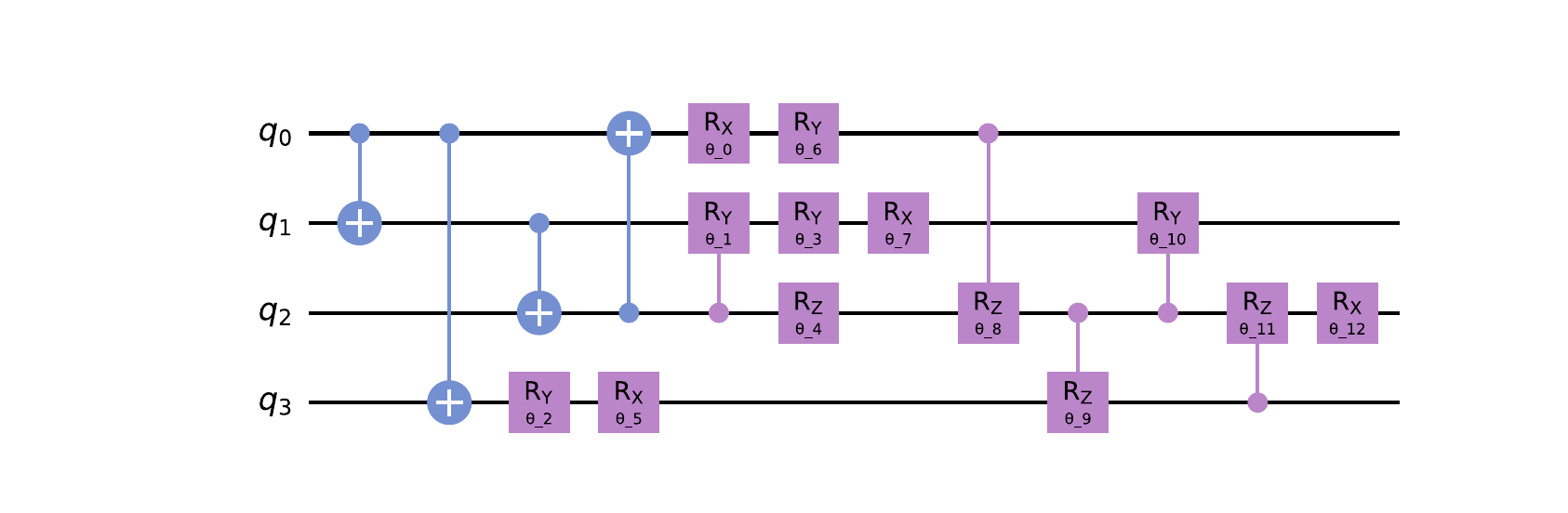}
\\

   (1.c)
\\
   \includegraphics[width=1\linewidth]{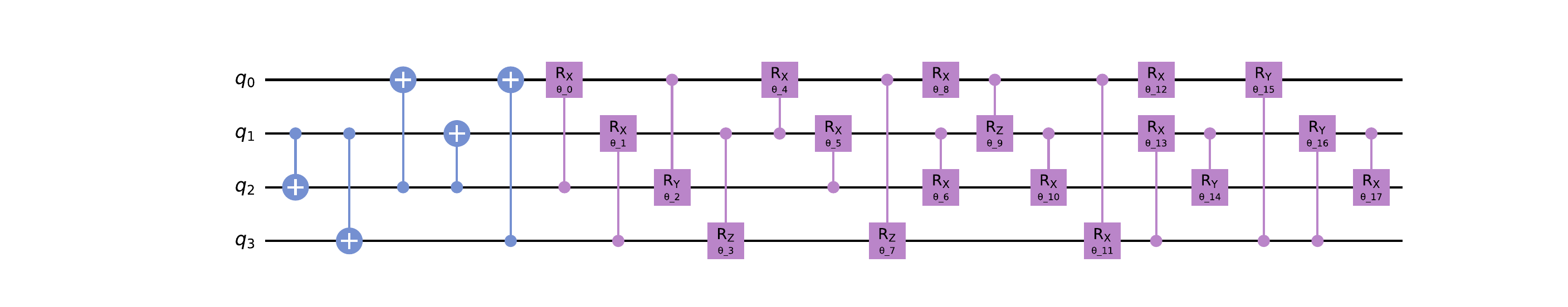}\\

   (2.a)
\\
   \includegraphics[width=0.8\linewidth]{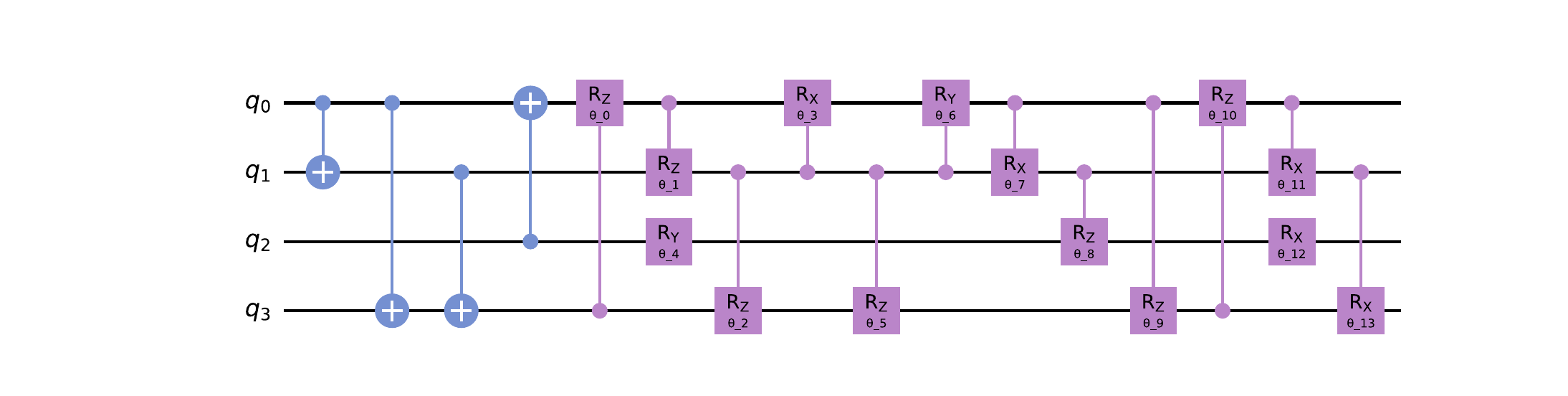}\\

   (2.b)
\\
   
   \includegraphics[width=0.75\linewidth]{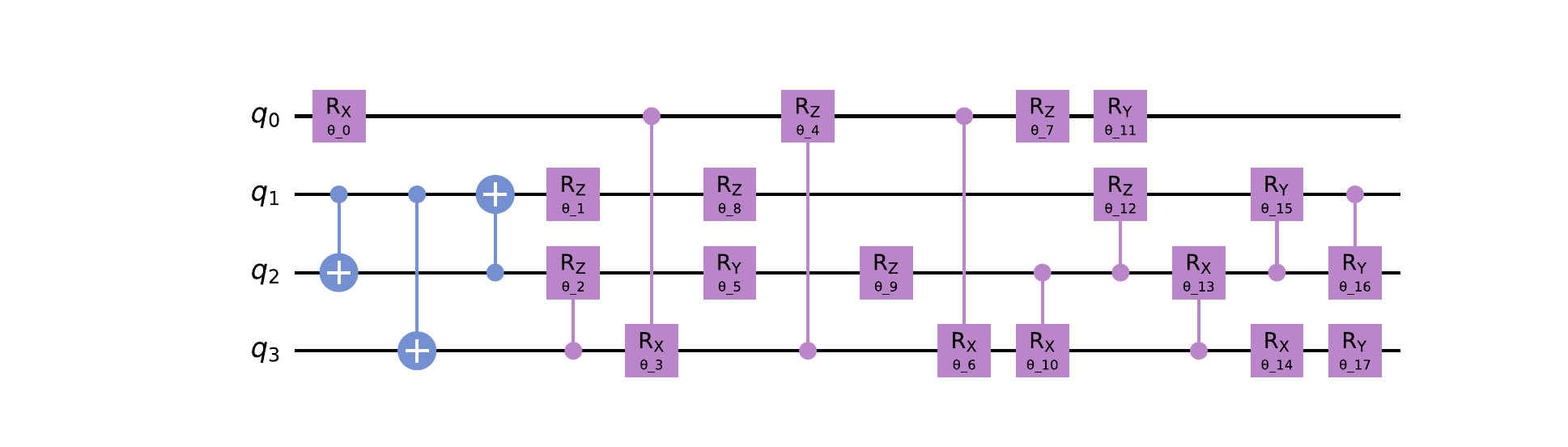}
\\

   (2.c)

\end{tabular}
\caption{Best BPQCO circuits in the (a) ideal environment, (b) noisy environment and (c) multi-objective method for the (1) Synthetic and (2) Iris datasets.}
\label{fig:best_circuits}
\end{figure}

Then, the first variant of BPQCO (subsection \ref{sec:bpqcon}) approach is applied in the noisy environment, denoted as BPQCO$_n$. For this purpose, $400$ and $500$ trials are run for the Synthetic and Iris datasets, respectively. After that, the circuits represented in Figure \ref{fig:best_circuits} are obtained as those with the best performance in the noisy environment. When compared with those obtained previously, it can be seen that the number of gates decreases in both datasets. It is worth noting the decrease in the use of controlled gates. In the Synthetic dataset, there has been a decrease from $23$ controlled gates in the ideal environment to $18$ in the noisy environment. As for the Iris dataset, the number of controlled gates has been reduced from $21$ to $16$. It has specially decreased the number of controlled rotations on the X axis, from $9$ to $4$. While the number of controlled rotations on the Z axis has increased from $3$ to $7$. In both cases, the number of CNOT gates used to add entanglement has also decreased.

\begin{figure}[t]
\centering
\begin{subfigure}{\textwidth}
   \includegraphics[width=1\linewidth]{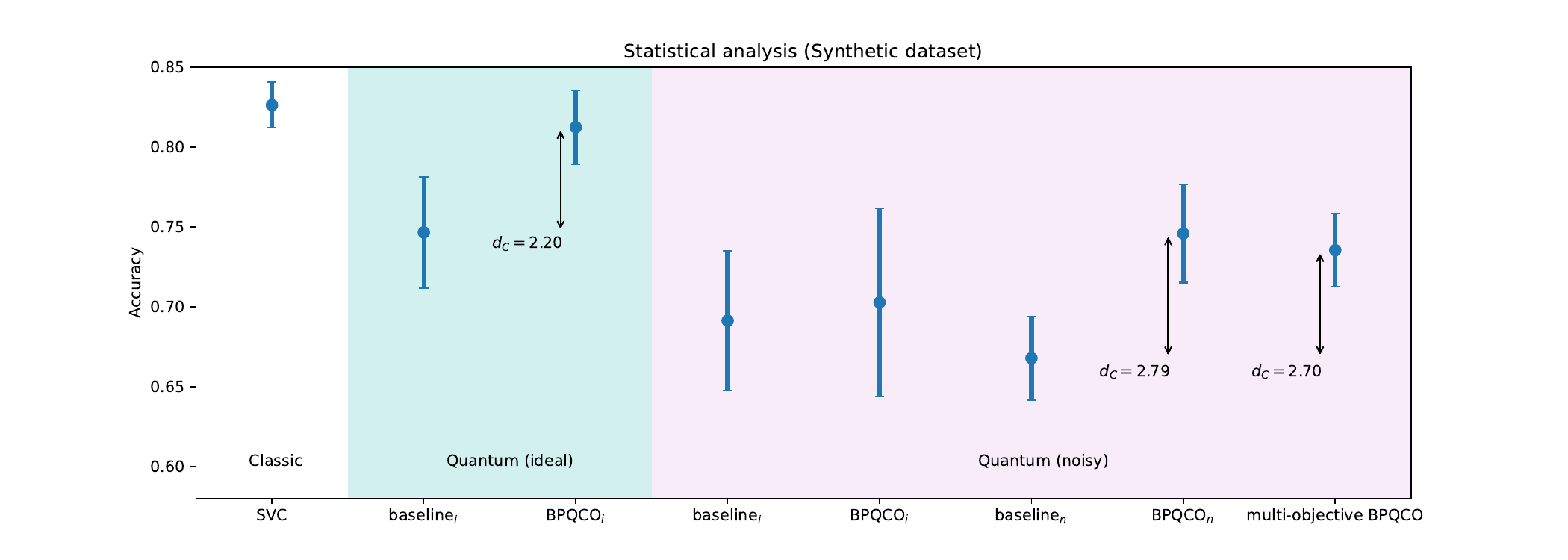}
   \caption{}
\end{subfigure}

\begin{subfigure}{\textwidth}
   \includegraphics[width=1\linewidth]{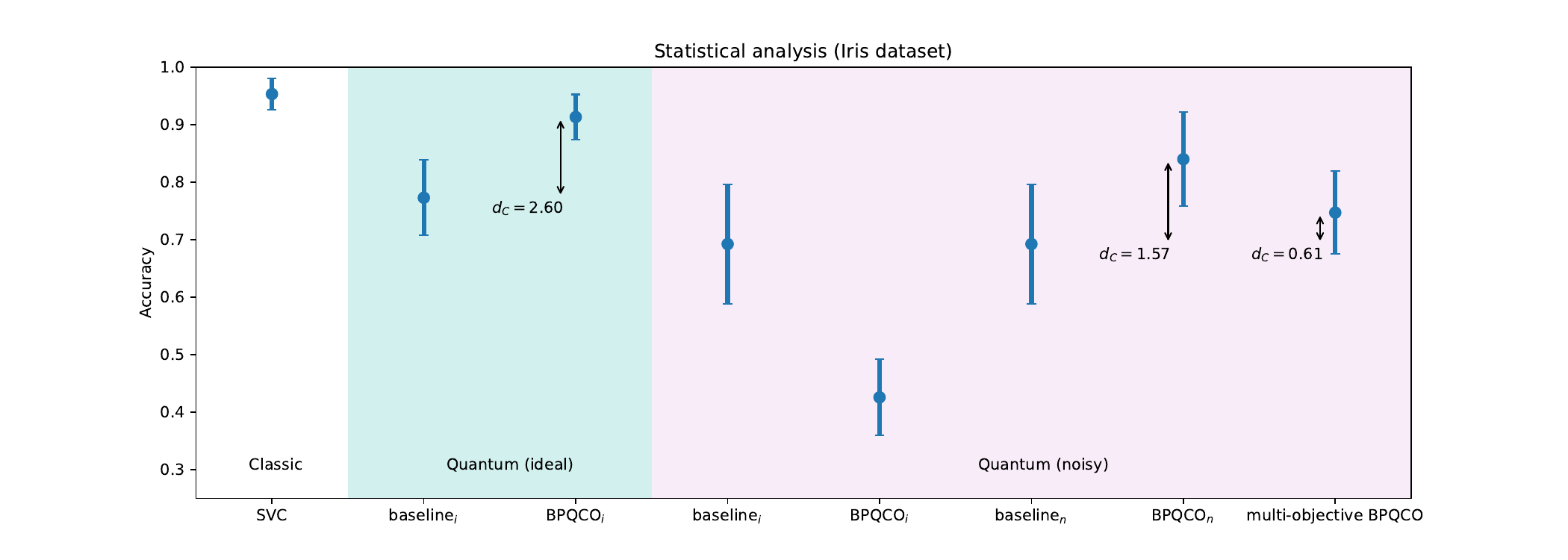}
   \caption{}
\end{subfigure}

\caption{Average Accuracy and standard deviation obtained for each design for the (a) Synthetic and (b) Iris datasets. The subscript indicates under which conditions ($i$ for ideal and $n$ for noisy) the circuit has been designed. The background shows in which environment each design has been run: white for the classical SVC, green for the ideal quantum environment and violet for the noisy quantum environment. Cohen's coefficient $d_C$ is calculated for each design obtained by BPQCO and its variants with respect to the baseline of each case (ideal or noisy).}
\label{fig:synthetic_iris_results}
\end{figure}

Finally, the multi-objective BPQCO (\ref{sec:mobpqco}) is applied, running the same number of trials as in the previous case for each dataset. It is reminded that this approach performs the optimization (design) of the circuit in the ideal environment with the objective of being subsequently executed in the noisy environment. In this case, there is no ``best'' circuit, but a set of efficient solutions, defined by a Pareto Front, is obtained. Therefore, this set of circuits is evaluated and the one with the best performance in terms of average accuracy and stability is selected. After that, the circuits shown in Figure \ref{fig:best_circuits} are selected. Analyzing the obtained circuits, it is evident that they are much simpler, greatly reducing the number of gates used. Again, the major reduction in both cases is in the number of controlled gates, especially the CNOTs for entanglement. Table \ref{tab:circuit_costs} shows the value of the complexity function associated with each circuit. Evidently, the lowest complexity is that of the multi-objective approach, since in this approach the aim was to minimize this aspect while maximizing performance.

The results on average ($10$-fold with $5$ seed initialization of each backend) of the whole complete process are presented in Figure \ref{fig:synthetic_iris_results}. In it, the potential of BPQCO in both the ideal and noisy environment can be observed. In addition, in order to complete the statistical study of the results, Cohen's coefficient  $d_C$ is used to quantify the effect size. It is considered that if the value is greater than $0.5$ the effect is medium and if it is bigger than $0.8$ the effect is large \citep{sawilowsky2009new}. Therefore, it can be concluded that the effect is large in the ideal environment and medium-large in the noisy environment.

\begin{table}[t]
	\caption{Complexity of each of the circuits obtained by the different BPQCO variants.}
	\centering
	\begin{tabular}{lll}
		\toprule
		& \multicolumn{2}{c}{Dataset}                   \\
		\cmidrule(r){2-3}
		Ansatz & Synthetic & Iris  \\
		\midrule
		BPQCO$_i$ & $0.5693$ & $0.4886$     \\
		BPQCO$_n$ & $0.3695$ & $0.3385$     \\
		Multi-objective BPCQCO  & $0.1854$ & $0.2205$    \\
		\bottomrule
	\end{tabular}
	\label{tab:circuit_costs}
\end{table}

\section{Conclusions and further work}\label{sec:conclusions}
Throughout this article, the importance of designing specific PQCs to solve a problem has been demonstrated. Of course, there are agnostic circuit templates that can provide good performance. However, this process of adapting the circuit involves a significant increase in the performance of the solution. For this purpose, we decided to use Bayesian optimization for circuit design, but other search strategies can be equally effective. To this end, we have presented different design aspects and a nomenclature/formulation with the aim of standardizing the parameters to be explored during the search. After extensive testing (analyzing not only performance but also stability), the BPQCO process has significantly improved the results obtained by predefined circuits.

Nonetheless, the most outstanding contribution of this work is the additional step taken to evaluate these circuits in real hardware conditions. This has shown the significant degradation that results presented in an ideal environment can suffer in today's noisy environments. To this end, two alternatives have been proposed, focusing on the adaptation of the circuit to the hardware. Again, the results obtained have met the expectations and objectives set, since they significantly improve the performance, resulting in circuits with less complexity and, therefore, less vulnerable to the effects of noise.

On the other hand, in order to simplify the experiments, the PQC design has focused on the search for the best ansatz. However, other aspects can be included in the search process, such as the number of qubits, the depth of the system, or the data encoding method, in which data re-uploading can be explored. In addition, the gates to be used can be modified by those supported by photonic computers for the use of continuous variable \citep{cv_nn}.

Finally, it is worth mentioning that we intend to incorporate the BPQCO process to other open lines. The goal is to adapt PQCs used for solving other more complex problems in the field of Earth Observation in which we are currently involved. This includes the design of hybrid models that incorporate VQCs in their architecture, as well as the design of VQEs for optimization tasks \citep{makarov2024quantum}.

\section*{Acknowledgments}\label{sec:ack}
This work has been supported by the Spanish Ministry of Science and Innovation under the Recovery, Transformation and Resilience Plan (Misiones CUCO Grant MIG-20211005).

\bibliographystyle{unsrtnat}
\bibliography{references}  

\end{document}